\documentclass{aa}
\usepackage[varg]{txfonts}
\usepackage{graphicx}
\usepackage{tablefootnote}
\usepackage{fixltx2e}

\begin{document}

\title{Possibility for albedo estimation of exomoons: Why should we care about M dwarfs?}

\author{Vera Dobos\inst{1}
\and \'Akos Kereszturi\inst{1}
\and Andr\'as P\'al\inst{1}
\and L\'aszl\'o L. Kiss\inst{1, 2}}

\titlerunning{Possibility for albedo estimation of exomoons}
\authorrunning{Dobos et al.}

\offprints{V. Dobos, \email{dobos@konkoly.hu}}

\institute{Konkoly Thege Mikl\'os Astronomical Institute, Research Centre for Astronomy and Earth Sciences, Hungarian Academy of Sciences, H-1121 Konkoly Thege Mikl\'os \'ut 15-17, Budapest, Hungary
  \and Sydney Institute for Astronomy, School of Physics A28, University of Sydney, NSW 2006, Australia}

\date{Received 1 June 2015 / Accepted 1 June 2015}

\abstract
{Occultation light curves of exomoons may give information on their albedo and hence indicate the presence of ice cover on the surface. Icy moons might have subsurface oceans thus these may potentially be habitable.
The objective of our paper is to determine whether next generation telescopes will be capable of albedo estimations for icy exomoons using their occultation light curves.
The  success of the measurements depends on the depth of the moon's occultation in the light curve and on the sensitivity of the used instruments. We applied simple calculations for different stellar masses in the V and J photometric bands, and compared the flux drop caused by the moon's occultation and the estimated photon noise of next generation missions with 5 $\sigma$ confidence.
We found that albedo estimation by this method is not feasible for moons of solar-like stars, but small M dwarfs are better candidates for such measurements. Our calculations in the J photometric band show that E-ELT MICADO's photon noise is just about 4 ppm greater than the flux difference caused by a 2 Earth-radii icy satellite in a circular orbit at the snowline of an 0.1 stellar mass star. However, considering only photon noise underestimates the real expected noise, because other noise sources, such as CCD read-out and dark signal become significant in the near infrared measurements. Hence we conclude that occultation measurements with next generation missions are far too challenging, even in the case of large, icy moons at the snowline of small M dwarfs. We also discuss the role of the parameters that were neglected in the calculations, e.g. inclination, eccentricity, orbiting direction of the moon.
We predict that the first albedo estimations of exomoons will probably be made for large icy moons around the snowline of M4 -- M9 type main sequence stars.}

\maketitle

\section{Introduction}

The Solar System provides useful analogues to extrapolate extrasolar satellite properties. The satellites of the giant planets hold substantial mass of water ice that is often located on the surface \citep{brown97, verbiscer90}. Although there could be differences between satellite systems, the icy surface, including water ice, is abundant on them \citep{stevenson85, sasaki10}. Analysing the optical properties of the Solar System satellites, it can be seen that the high albedo in general indicates water ice. Non-ice covered satellites usually show low geometric albedo values between 0.04 and 0.15, while ice covered satellites show higher values, although different ingredients (mostly silicate grains) could decrease the albedo if they are embedded in the ice, or the ice is old and strongly irradiated/solar gardened \citep{grundy00, sheppard05}. These satellites show moderate geometric albedo between 0.21 for Umbriel and 0.99 for Dione \citep{veverka91, verbiscer13}. Beside water ice, nitrogen ice is also a strong reflector, producing albedo around 0.76 on Triton \citep{hicks04}. On this satellite the bright surface can partly be explained by its relatively young, 6-10 Myr old age. \citep{schenk07}. The most important analogues for high albedo in the Solar System could be Europa with 1.02 and Enceladus with 1.38 geometric albedo \citep{verbiscer13}. In the case of Europa the relatively young (about 50 million year old) surface accompanies with clean and bright ice, while on Enceladus freshly fallen ice crystals increase the albedo significantly \citep{pappalardo99, porco06}. Based on these observations, we assume that geometric albedo above around 0.7 indicates some ice, often water ice, with relatively young surface, and higher albedo might suggest freshly fallen water ice crystals. In this work we are analysing exomoons with such high albedo, as it might give insight into the surface properties.

For exomoon detection, current techniques mostly focus on light curve analysis \citep[and references therein]{kipping09a, kipping09b, kippingetal09}. The concept is that the exomoon alters the light curve in a specific and identifiable way \citep{szabo06, simon07}. Other methods also have been proposed, such as direct imaging \citep{peters13}, radial velocity measurements \citep{simon10}, microlensing \citep{liebig10, bennett14}, or pulsar timing \citep{lewis08}. In this work we discuss how the light curves of occultations can provide insights for albedo estimation.

Habitability is also in the focus of exomoon researches. \citet{heller13} developed a model for calculating the incident flux on an exomoon arriving from stellar irradiation, reflective light from the planet, thermal radiation of the planet and from tidal heating of the moon. They also considered eclipses as possible energy sinks. Too much incident flux on a moon can trigger the runaway greenhouse effect, which leads to the loss of water reservoirs on the body, hence making it uninhabitable \citep{kasting88}. \citet{forgan13} developed a climate model for exomoons in order to investigate the energy balance of satellites. This model combined with eclipses and the ice-albedo feedback is capable of predicting the orbital parameters of a moon that will result in a snowball state, defining an outer limit for the exomoon habitable zone \citep{forgan14, forgan16}.

M dwarfs could have specific characteristics regarding the exomoon identification and analysis. Although most search projects focus on solar-like stars, M dwarfs as targets should also be considered, because of more than 70\% of the stars are of M spectral type. In theory it is possible that no large exomoons are present around M dwarfs, as their smaller stellar mass allow the formation of only smaller exoplanets and exomoons than in the Solar System. But based on the discovered exoplanets currently 68 M dwarfs are known to have 96 exoplanets, from which the heaviest planet's mass is 62 $M_\mathrm{J}$, 20 other exoplanets have at least 10 Jupiter-masses, and there are 13 other planets that are heavier than 1 $M_\mathrm{J}$ (source: \textit{exoplanet.eu}, January 2016). The average mass of the exoplanets of the 68 M dwarfs is 4.8 $M_\mathrm{J}$.

The mass of a moon is limited by the mass of its host planet \citep{crida12} and the mass of satellite systems is usually proportional to the mass of their host planet. \citet{canup06} showed that this might be the case for extrasolar satellite systems as well, giving an upper limit for the mass ratio at around 10\textsuperscript{$-4$}. Thus 10 Jupiter-mass planets may have 0.3 Earth-mass satellites. However, if the moon is not originated from the circumplanetary disk, but from collision, like in the case of the Earth's Moon, then even larger satellites might exist.

Moons are more likely to form by ejection around Neptune than around Jupiter-like giants because. Jupiter-like gas giants are composed of mostly liquid and gaseous hydrogen at their outer layers, thus the ejected satellite-forming material might be $\mathrm{H_2 O}$ poor, composed mostly of gaseous $\mathrm{H_2}$ and could easily escape. But if this event happens before the dispersion of the circumplanetary disk, the co-accreted ice in the disk could provide icy surface for such satellite. Neptune-like planets seem to be better candidates. They are mainly composed of $\mathrm{H_2 O}$, and the ejected satellite, too. The occurrence of impacts at giant planets are favoured by certain theories, as a contributor to the continuous gas accretion \citep{broeg12}, increasing the probability of impact-ejected icy satellites.

Besides ejection, capture could also produce a large moon. In the so-called 'binary-exchange' captures, a terrestrial sized moon can be captured by a 5 Jupiter-radii planet around an M dwarf, if the planetary encounter is close enough \citep{williams13}. For Neptune-mass planets the capture is easier than for Jupiter-mass planets, because generally the encounter speeds are lower. The capturing is much more difficult at small stellar distances, but if it occurs before or during the planetary migration, then it is possible that such a planet-moon pair will orbit an M dwarf at a close distance (around the snowline). In the Solar System, Triton may be an example for such binary-exchange capture \citep{agnor06}. For these reasons we consider the existence of large exomoons (up to a few Earth-masses) of Neptune-mass planets around M dwarfs.

The aim of this work is to propose a new method to identify elevated albedo and thus the possibility of water ice on the surface of an exomoon using photometric measurements during occultations. We also discuss the role of different orbital and physical parameters that may influence the observed data. As in the case of exomoons, surface ice cover might be more frequent than for exoplanets, based on examples of the Solar System and theoretical argumentation. Although the photometric signal would be small, it is worth considering the possibility and potential results of such measurements, because of the following reasons: the findings might orient the instrumental development, and search programs to focus on given stellar type, and also give a hint on surface properties of exomoons, or provide information on their habitability. Analysing different configurations helps to see the best possibility to estimate exomoons' albedo. We also aim to compare the scale and possible role of such configurations of exomoons that are realistic but not present in the Solar System. Just like in the case of exoplanets, unusual situations could be present and occasionally dominate because of observational selection effects (see for example the case of many hot Jupiters during the first decade of exoplanet discoveries). In this work we analyse these possibilities, pointing to the difficulties, possible ways and best conditions for such estimation.

Although such precise measurements are difficult to make, the method could be used to differentiate among exomoon surface properties. Despite no exomoon has been discovered yet, probably those are numerous in other planetary systems, and will play an important role in astrobiology research in the future \citep{kaltenegger10, heller13} giving rationality to elucidate the potential method outlined here. Because of the large number of possible configurations and different parameters of exomoon systems, specific cases are considered in this paper with the aim to give general insights about the factors that are important for albedo estimation. We also consider such possible conditions that are not present in the Solar System but could be observed at certain exomoons, e.g. large exomoons, objects on eccentric, inclined or retrograde orbits (Section \ref{diffpar}).

The structure of this paper is as follows. In Section 2 we describe the numerical treatment to determine the albedo of an exomoon and the photon noise level of different instruments. In Section 3 we present the results and consider both Solar System objects and hypothetical moons to imply our calculations. In Section 4 we discuss the rationality and limitations of this method which is followed by a Conclusion section.

\section{Methods}

In this section we present the background of albedo calculations from occultation light curve. Because of the wide range of possible conditions (distance of the moon from its host planet and host star, orbital elements, size, albedo etc.) only general and simple cases are used here. With this approach we get insight into the most important parameters without analysing all possibilities. The potential significance of those parameters that are not used in this model are discussed in Section \ref{diffpar}.

Three different configurations are possible just before and after any occultation:
\begin{enumerate}
\item the exomoon is behind the planet (both the planet and the moon are in the same line of sight),
\item the exomoon is in front of the planet (both the planet and the moon are in \ the same line of sight),
\item the exomoon and the planet are separately 'visible' (although spatially not resolved).
\end{enumerate}

It can be generally assumed that the configuration does not change much in most cases during the occultation. The reason is that the time duration of the occultation varies in a couple of hours time scale, while the orbital period of the exomoon is assumed to be longer \citep{cabrera07}. A reverse scenario is unlikely, but also discussed by \citet{simon10} and \citet{sato10}.

\begin{figure*}
	\centering
	\includegraphics[width=120mm]{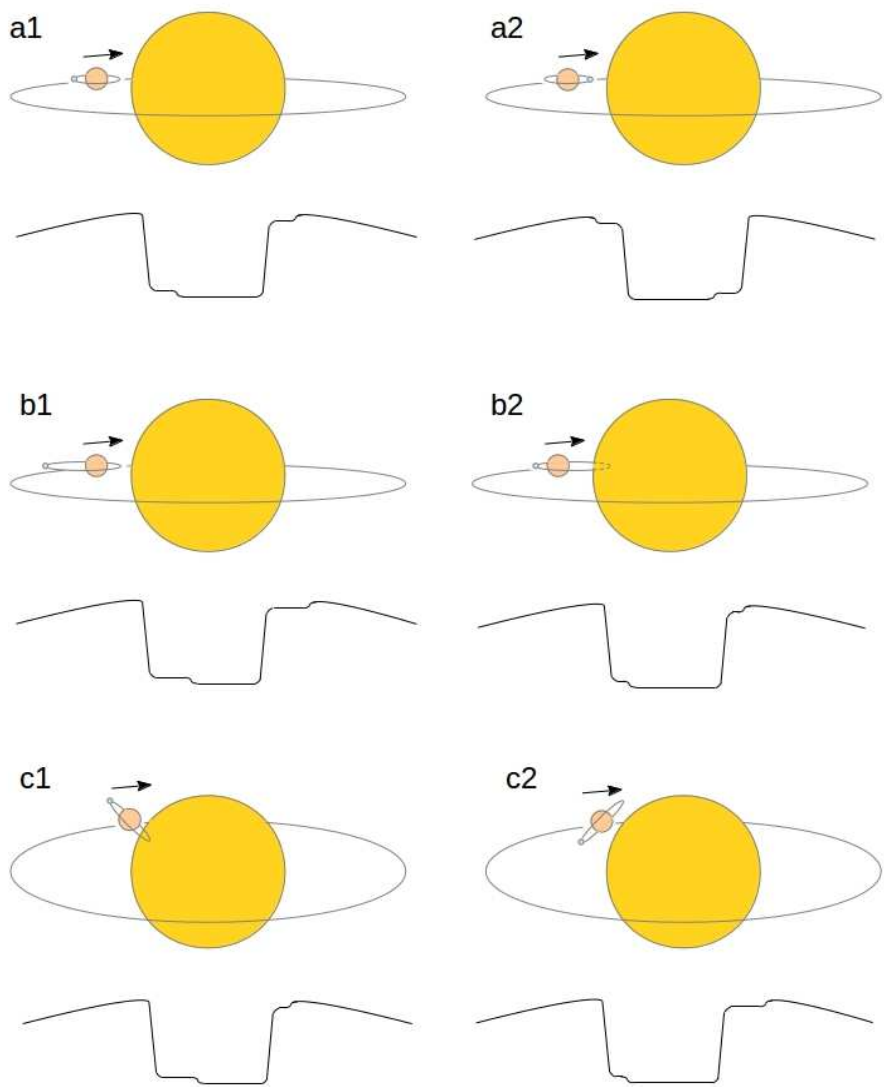}
	\caption{Schematic light curve during an occultation with an exomoon orbiting its exoplanet with some different basic conditions: a1 -- exomoon occults after the exoplanet, a2 -- exomoon occults before the exoplanet, b1 -- observationally favourable case of large eccentricity increasing the plateau duration, b2 -- unfavourable case of large eccentricity decreases the plateau duration, c1 -- favourable case of large inclination (relative to the plane perpendicular to the line of sight), c2 -- unfavourable case of large inclination. The sizes of the objects are not to scale.}
	\label{transit_types}
\end{figure*}

From the three possible configurations, the third (3) is the one, which favours for the identification of the exomoon, since in the other cases the moon's presence does not change the shape of the light curve (although the depth of the flux minimum can be different). The third configuration is also the most probable of all, based on geometric considerations \citep{heller14a}. Apart from a few special cases, the moon spends just a small fraction of its orbital time in front of, or behind the planet \citep{heller12}. In this (third) case two sub-configurations are possible: the planet occults before the moon (a1 in Fig. \ref{transit_types}), or the moon occults before the planet (a2 in Fig. \ref{transit_types}). These cases can be seen in Figure 1. along with schematic light curves. Note that unlike at transits, stellar limb darkening does not affect the light curve at occultations.

\subsection{Albedo}

\begin{figure}
	\centering
	\includegraphics[width=80mm]{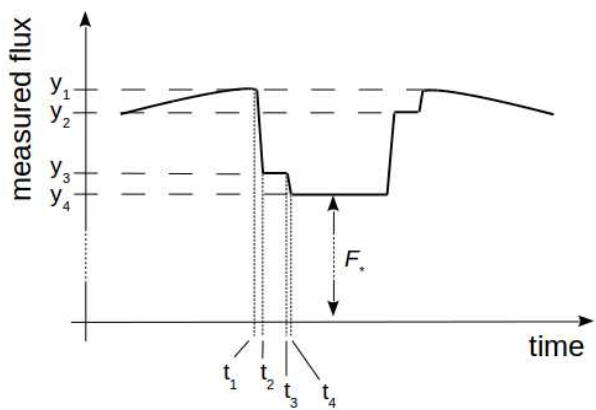}
	\caption{Schematic representation of the light curve \textup{(case a1 in Fig. \ref{transit_types})} during the occultation of an exoplanet with an exomoon.}
	\label{light_curve}
\end{figure}

We use simple formulae for the calculations, because this is a basic \ phenomenon of exomoons that has not been analysed before. Simplified cases are considered because of the large number of possibilities and also to see the big picture without being lost in the details.

For the albedo calculation we considered the following formulae and configuration. Fig. \ref{light_curve} shows a schematic light curve of an occultation, assuming that first the planet moves behind the star, followed by the moon (case a1 in Fig. \ref{transit_types}). The light curve has its maximum at $y_1$ and the minimum of the occultation is at $y_4$. The latter corresponds to the stellar flux (the planet and the moon are behind the star), denoted by $F_\star$ which is normalized so that $F_\star = 1$. Reflected stellar light from the planet and from the moon give contributions to the observed flux. Values of $y_3$ and $y_2$ refer to the measured flux before the occultation of the moon and after the occultation of the planet (the moon is still occulting), respectively. Measuring both $y_1 - y_2$ and $y_3 - y_4$ can be used to refine their value (which should be equal in theory), assuming that the apparent configuration of the system does not significantly change during the occultation. The accuracy of this value is very important, because it plays a key role in calculating the moon's albedo.

For the difference of the measured fluxes:
\begin{equation}
	y_3 - y_4 = \frac{F_\mathrm{m}} {F_\star} \, ,
\end{equation}

\noindent where $F_\mathrm{m}/F_\star$ is the relative flux contribution of the moon normalized to the stellar flux. The flux contribution of the planet is irrelevant for our study. We call the $y_3 - y_4$ difference moon occultation (MO) depth. Adapting Eq. (2) from \citet{rowe08} to this case, the ratio of the fluxes can also be written as 
\begin{equation}
	\frac{F_\mathrm{m}} {F_\star} = A_\mathrm{g} \left( \frac{R_\mathrm{m}} {a_\mathrm{p}} \right)^2 \, ,
\end{equation}

\noindent where $A_\mathrm{g}$ is the geometric albedo, $R_\mathrm{m}$ is the radius and $a_\mathrm{p}$ is the distance of the exomoon from its host star, which is approximated by the semi-major axis of the planet. The radius of the moon is assumed to be already known from previous transit observations. The geometric albedo is the ratio of brightness at zero phase angle compared to an idealized, flat, fully reflecting, diffusively scattering disk. The Bond albedo on the other hand, is the reflectivity at all phase angles. The value of the Bond albedo is restricted between 0 and 1. The geometric albedo on the other hand, can occasionally be higher than 1, especially for atmosphereless bodies with icy surface (e. g. Enceladus, Tethys) because of the strong backscattering and opposition effect \citep{verbiscer07}. Because of the coherent backscatter opposition effect (CBOE), the phase difference of the lightwaves in the exact backscatter direction will be zero and the amplitudes add coherently as they interfere with each other, resulting in higher intensity up to a factor of two \citep{hapke12, verbiscer13}.

From Eqs. (1) and (2) one can get:
\begin{equation}
	A_\mathrm{g} = \left( y_3 - y_4 \right) \left( \frac {a_\mathrm{p}} {R_\mathrm{m}} \right)^2 \, .
\end{equation}

The geometric albedo of an exomoon can be calculated by obtaining the MO depth from observations during occultation, and by using Eq. (3). However, very precise measurements are required, hence it is reasonable to consider the MO depth as an upper limit \citep{rowe06}. Several observations are needed for a phase-folded light curve that show the presence of the exomoon because of the photometric orbital sampling effect \citep{heller14a}.

 Using Eq. (3) we calculated the MO depth values for different bodies, in order to investigate the required precision of future observations. These photometric measurements may lead to obtain the albedo of an exomoon with the use of the described method.

\subsection{Calculation of 'plateau duration' and 'MO duration'}

In order to achieve the detection of small flux differences, long enough detector integration time is required. An important limiting factor for detection is the duration while the exomoon is 'visible' alone and the exoplanet is behind the star (the time interval $t_3 - t_2$ in Fig. \ref{light_curve}), or when only the exoplanet is visible without the moon. We call this phase 'plateau duration', because it causes a plateau in the light curve.

The plateau duration is calculated for ideal cases, when the exomoon is located at its maximum elongation from the exoplanet (where it has only radial velocity component). Circular orbits were assumed both for the planet and for the moon. This is a crude simplification, but in case of captured or impact-ejected moons (like Triton around Neptune or the Moon around the Earth) tidal interactions can strongly reduce the satellite's eccentricity, almost circularizing the orbit. With this simplified model the plateau duration can easily be calculated by using the speed of the planet-moon pair around the central star:
\begin{equation}
	v_\mathrm{p} = \sqrt{ \frac{G \left( M_\star + M_\mathrm{p} \right) } {a_\mathrm{p}} } \, ,
\end{equation}

\noindent where $v_\mathrm{p}$ is the orbital speed of the planet, $G$ is the gravitational constant, $M_\star$ and $M_\mathrm{p}$ are the masses of the star and the planet, respectively. Since circular orbits are considered, $a_\mathrm{p}$ indicates the planet's orbital distance. In most cases the apparent location of the moon does not change significantly during the occultation. For this reason the plateau duration can be estimated from the orbital speed of the planet, if the orbital distance of the moon ($a_\mathrm{m}$) is known (this is the distance that the moon needs to travel after the planet disappears behind the star). If the moon was not in its maximal elongation, then the distance (and hence the plateau duration) would be shorter.

However, the moon can be fast enough on its own orbit around the planet to influence the plateau duration. In case the moon passes 25\% of its orbit (starting from its maximal elongation) then we terminate the plateau duration, because the moon must be in front of or behind the planet. These cases were taken into account as well. The orbital speed of the moon ($v_\mathrm{m}$) was calculated by
\begin{equation}
	v_\mathrm{m} = \sqrt{ \frac{G M_\mathrm{p}} {a_\mathrm{m}} } \, .
\end{equation}

\noindent Both time intervals were calculated: $t_\mathrm{a} = a_\mathrm{m} / v_\mathrm{p}$ and $t_{25} = 0.25 \left(2 a_\mathrm{m} \pi \right) / v_\mathrm{m}$, and the shorter was considered as the plateau duration ($T_{event}$) in each case.

The plateau duration was calculated for different mass host planets. It can be shown that having 1, 5, or 10 $M_\mathrm{J}$ mass planets ($M_\mathrm{J}$ denotes the mass of Jupiter) does not change the order of magnitude of the results. Table \ref{MJ} shows such cases for two moons. The stellar distance is 3 AU in each case, and a Sun-like star was used in the calculations. For the calculation, the radius of the planet was calculated from the mean density (1326 kg/m\textsuperscript{3}) and original radius (71,492 km) of Jupiter in each case. For the orbital distance of the two moons the real semi-major axes were used.

\begin{table*}
	\caption{Plateau duration for different planetary masses at 3 AU distance from a Sun-like star.}
    \label{MJ}
	\centering
	\begin{tabular}{c c c c c}
		\hline\hline
		 & & plateau duration & plateau duration & plateau duration \\
		Name & $a$ [km] & with a 1 $M_\mathrm{J}$ mass & with a 5 $M_\mathrm{J}$ mass & with a 10 $M_\mathrm{J}$ mass \\
		 & & planet [h] & planet [h] & planet [h] \\
		\hline
		Enceladus & 237,378 & 3.83 & 2.00 & 1.42 \\
		Europa & 670,900 & 10.83 & 9.52 & 6.74 \\
		\hline
	\end{tabular}
\end{table*}

The plateau duration depends on many factors, including the exomoon's apparent distance from its host planet, its orbital velocity and the size of the host star. Larger stellar disk, lower velocity and larger planetary distance of the moon increase the length of the plateau duration. If the apparent planet-moon distance is larger than the diameter of the star, then the 'plateau' will not appear in the light curve, because the planet and the moon will occult separately, making two separate drops in the light curve (i.e. the planet's occultation ends before the moon's occultation starts or the moon's occultation ends before the planet's starts). In other words the occultations of the moon and the planet will not overlap in time. In this case the plateau duration is replaced by the duration of the moon's occultation. We call it 'moon occultation duration', or MO duration for short.

In these cases, when the size of the star limits the MO duration, the following simple calculation was used. The velocity of the moon is already known from previous calculations, and the distance that the moon travels during the MO duration approximately equals to the diameter of the star. Since it is assumed that the impact parameter of the moon and the inclination of both the planet and the moon are zeros, the related time can be calculated from $t_\mathrm{MO} = 2 R_\star / v_\mathrm{p}$. If $t_\mathrm{MO} < t_\mathrm{a}$, then $t_\mathrm{MO}$ will be considered as the MO duration ($T_{event}$).

%\begin{figure}
%	\centering
%	\includegraphics[width=80mm]{inklinacio_magyarazat_peldak_Neptunusszal.eps}
%	\caption{Plateau / MO duration related distances (as black horizontal bars) at different sized central stars with the same exomoon orbital distance (equal to that of Europa around Jupiter) but at different inclinations ($0^{\circ}$, $30^{\circ}$, $60^{\circ}$, $90^{\circ}$). a) Solar sized star, b) M0 and c) M9 main sequence star sizes are indicated.}
%end{figure}

\begin{figure}
	\centering
	\includegraphics[width=80mm]{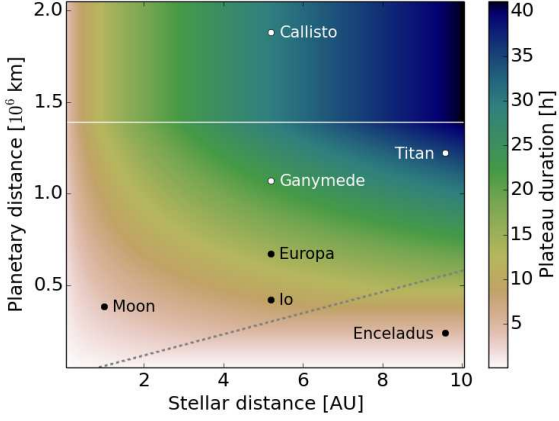}
	\caption{The so-called 'plateau duration' for different stellar and planetary distances. Examples from the Solar System are plotted for comparison.}
	\label{plateau}
\end{figure}

The phase space of different stellar and planetary distances of a hypothetical moon was mapped. The plateau duration for different stellar and planetary distances can be seen in Fig. \ref{plateau}. The masses of the star and the planet are 1 solar mass and 1$M_\mathrm{J}$, respectively, for all cases. Black and white colours indicate the longest (41 hours) and the shortest (6 minutes) durations, respectively. Below the dashed grey line the plateau duration is terminated, because the moon reaches 25\% of its orbit. It mainly occurs for higher stellar distances, which produces lower orbital velocity around the star. Because of this lower velocity $t_{25}$ will be smaller than $t_\mathrm{a}$. Above the solid horizontal white line the plateau duration is constant for each stellar distance. The growing semi-major axis of the moon results in larger $t_\mathrm{a}$, but at about 1,393,000~km ($2 R_\star$) $t_\mathrm{a} = t_\mathrm{MO}$, or in other words, above this planetary distance the planet and the moon occults separately (the MO duration takes the place of the plateau duration). A few examples from the Solar System are plotted for comparison.

\subsection{Calculating the photon noise level of different instruments} \label{calcphotonnoise}

Let us assume that the star emits perfect black-body radiation. This is 
a good assumption if we intend to estimate the photon noise level 
for a given instrument. The star is characterized by its radius $R_\star$ and 
$T_\star$, the effective  temperature of the stellar surface.
According to Planck's law, the power density on unit surface and 
unit solid angle as the function of the wavelength $\lambda$ can be written as
\begin{equation}
B_\lambda(\lambda){\rm d}\lambda=\frac{2hc^2}{\lambda^5}\frac{1}{e^{\lambda_\star/\lambda}-1}{\rm d}\lambda\label{eq:blambda}
\end{equation}

\noindent where
\begin{equation}
\lambda_\star=\frac{hc}{k_{\rm B}T_\star},
\end{equation}

\noindent which is proportional to the location of the peak of the spectral 
energy distribution. The constants are the following: $k_{\rm B}$ is the Boltzmann constant,
$h$ is the Planck constant and $c$ is the speed of light.
The integral over the stellar surface $(4\pi R_\star^2)$ as well as on the 
irradiated hemisphere according to Lambert's cosine law yields a total power 
output at a certain wavelength which can be computed as
\begin{equation}
P_\lambda(\lambda){\rm d}\lambda=4\pi^2 R_\star^2B_\lambda(\lambda){\rm d}\lambda.
\end{equation}

The number of photons emitted at a certain wavelength can then be computed
by dividing the power $P_\lambda(\lambda)$ by the photon energy $hc/\lambda$.
The total number of photons detected by a detector having a gross
quantum efficiency of $Q(\lambda)$ at unit time is then
\begin{equation}
\frac{{\rm d}n}{{\rm d}t}=
\frac{A}{4\pi d^2}
\int\limits_{\lambda=0}^\infty
Q(\lambda)P_\lambda(\lambda)\left(\frac{hc}{\lambda}\right)^{-1}{\rm d}\lambda
\end{equation}

\noindent Here $A$ is the area of the entrance aperture on the telescope 
(on which the detector is mounted) and $d$ is the distance to the star. 
In practical applications, $Q(\lambda)$ is the product of 
the transparencies of various filters, lens and reflective elements 
forming the telescope optics and the quantum efficiency of the 
detector itself. In order to simplify our calculations, $Q(\lambda)$ can be 
written as a characteristic function having a constant value of $Q$ within an interval
of $[\lambda_{\rm min},\lambda_{\rm max}]$ (and it is zero outside this interval).
Let us define $\Delta\lambda$ as the effective bandwidth of the filter as 
$\Delta\lambda=\lambda_{\rm max}-\lambda_{\rm min}$ and the central
wavelength as $\lambda_{\rm f}=(\lambda_{\rm min}+\lambda_{\rm max})/2$.

Using this assumption above, the total number of photons received at unit time
can be written as
\begin{equation}
\frac{{\rm d}n}{{\rm d}t}=
2\pi AQc\left(\frac{R_\star}{d}\right)^2
\int\limits_{\lambda=\lambda_{\rm min}}^{\lambda_{\rm max}}
\frac{1}{\lambda^4}\frac{1}{e^{\lambda_\star/\lambda}-1}{\rm d}\lambda
\end{equation}

\noindent By introducing the variable $x=\lambda_\star/\lambda$ and applying 
the respective measure of ${\rm d}\lambda=(\lambda_\star/x^2){\rm d}x$,
we can write
\begin{equation}
\frac{{\rm d}n}{{\rm d}t}=
2\pi A\left(\frac{R_\star}{d}\right)^2
c\left(\frac{k_{\rm B}T_\star}{hc}\right)^3
Q\int\limits_{x=\lambda_\star/\lambda_{\rm max}}^{\lambda_\star/\lambda_{\rm min}}
\frac{x^2}{e^x-1}{\rm d}x\label{eq:photoncountintegral}
\end{equation}

\noindent If the bandwidth of the filter is narrow, i.e. $\Delta\lambda\ll\lambda_{\rm f}$,
the integral above can further be simplified by assuming the
expression $x^2(e^x-1)^{-1}$ to be constant in the 
interval of $[\lambda_\star/\lambda_{\rm max},\lambda_\star/\lambda_{\rm min}]$.
In this case, $\Delta x=x_{\rm max}-x_{\rm min}$ can be approximated
as $\Delta x\approx\lambda_\star\Delta\lambda/\lambda_{\rm f}^2$ and hence
the photon flow is
\begin{equation}
\frac{{\rm d}n}{{\rm d}t}=
2\pi A\left(\frac{R_\star}{d}\right)^2
Qc\left(\frac{\Delta\lambda}{\lambda_{\rm f}}\right)\left(\frac{1}{\lambda_{\rm f}^3}\right)\frac{1}{e^{\lambda_\star/\lambda_{\rm f}}-1}\label{eq:pnoise}
\end{equation}

\noindent Once the photon flow, i.e. the number of photons received by the detector
is known, the photon noise of the instrument can be computed for any 
exposure time ($T$) as
\begin{equation}
\frac{\Delta n}{n}=\left(T\cdot\frac{{\rm d}n}{{\rm d}t}\right)^{-1/2}.
\end{equation}

If the filter bandwidth is comparable to the central wavelength,
the integral in Eq.~(\ref{eq:photoncountintegral}) can be computed numerically.
Since the integrand is a well-behaved function, even a few steps using
Simpson's rule can be an efficient numerical method for this computation.
We note that in the case of Kepler or TESS or PLATO 2.0, where
CCDs are used without any filters, the bandwidth of the quantum efficiency
curve is comparable to its central wavelength, so this numerical 
computation is more accurate. However, the error in the accuracy introduced
by this simplification is comparable to the discretization of the 
$Q(\lambda)$ function (i.e. it is in the range of $10-20\%$ in total,
depending on the actual values of $\lambda_\star$, $\lambda_{\rm f}$ 
and $\Delta\lambda$).

In order to apply Eq.~(\ref{eq:pnoise}) in practice, it is worth converting
$R_\star$, $d$, $\lambda_{\rm f}$ and $\lambda_\star$ in astronomically relevant units 
(such as solar radius, parsec and microns) instead of SI units. After these
substitutions, we can write the approximation
\begin{equation}
\frac{{\rm d}n}{{\rm d}t}\cong
10^{12}\,{\rm Hz}
\left(\frac{A}{\rm m^2}\right)
\left(\frac{R_\star/R_\odot}{d/{\rm pc}}\right)^2
Q
\left(\frac{\Delta\lambda}{\lambda_{\rm f}}\right)
\frac{1}{(\lambda_{\rm f}/{\rm\mu m})^3}
\frac{1}{e^{\lambda_\star/\lambda_{\rm f}}-1}.\label{eq:pnoise2}
\end{equation}

\noindent This approximation is accurate within a few percents w.r.t.
Eq.~(\ref{eq:pnoise}). In the constant $10^{12}\,{\rm Hz}$ we included
both the dimension conversions (solar radius, parsec, microns) as well 
as the other constants such as $2\pi$ and the speed of light.

If we intend to detect a signal $S$ over the (cumulative) time $T$, 
the corresponding signal-to-noise ratio is going to be
\begin{equation}
{\rm S/N}=S\cdot\left(T\cdot\frac{{\rm d}n}{{\rm d}t}\right)^{1/2}.
\end{equation} 

\noindent The integration time is calculated as the product of the MO duration (or plateau duration) and the number of events (occultations): $T=T_{event} \cdot N_{event}$. For observatories, 30 events are assumed for the measurements, and for survey missions the number of events is calculated from the length of the mission campaign and the orbital period of the planet: $N_{event}=T_{campaign}/P_p$.

For all the calculations presented in this work, $d=50$ pc and $S/N=5$ is used, the latter corresponds to $5\sigma$ detection. In some cases (see Table \ref{instruments} for specific cases), instead of the $Q$ quantum efficiency, the system throughput was used, which contains $Q$ and other optical attenuations as well. In order to make these different cases comparable, either the system throughputs, or 80 percent of the quantum efficiency ($0.8Q$) was used in the calculations.

%We present the results of the calculations in the following section.

\subsection{Orbital distances around M dwarfs}

Beside solar-like stars, the calculations are applied to a range of small M dwarfs. The orbital distances of the planet and the moon are calculated as functions of the stellar mass. In case of M dwarf host stars, the planet-moon pair is set to the snowline. Icy exomoons are expected to form beyond or around the so-called snowline, which is the distance where water ice condenses in the protoplanetary disk, supporting the formation of large mass planetesimals that finally evolve toward gravitationally collected gas giants \citep{ida08, qi13}.

The location of the snowline is calculated from the equilibrium temperature at the planet's sub-stellar point ($T_0$)  \citep{cowan11}: 
\begin{equation}
	a_\mathrm{p} = T_\star^2 \frac{R_\star} {T_0^2} \, .
	\label{snowline}
\end{equation}

\noindent $T_0$ is assumed to be approximately 230 K at the ice condensation boundary. To obtain $R_\star$ and $T_\star$ from the stellar mass ($M_\star$), a parabolic equation was fitted to the effective temperature and radius values given by \citet[Table 1]{kaltenegger09}. Only M4 -- M9 stars were used for the fitting. The obtained equations are:
\begin{equation}
	R_{\star} = -0.12 + 3.31 M_\star - 7.12 M_\star^2 \, ,
    \label{R_star}
\end{equation}
%\noindent and
\begin{equation}
	T_{\star} = 1496 + 13301 M_\star - 26603 M_\star^2 \, ,
    \label{T_star}
\end{equation}

\noindent that can be used if $0.075 < M_\star < 0.2$.

The moon's orbital distance is set to 0.45 Hill radius ($R_\mathrm{H}$), hence it also depends on the stellar mass. It was calculated from the following formula:
\begin{equation}
	a_\mathrm{m} = 0.45 \, R_\mathrm{H} = 0.45 \, a_\mathrm{p} \left( \frac{M_\mathrm{p}} {3 M_\star} \right)^{1/3} \, .
    \label{M_star}
\end{equation}

\noindent We chose this distance in order to avoid orbital instability, since \citet{domingos06} and \citet{donnis10} showed that beyond approximately half of the Hill radius, the orbit of a direct orbiting exomoon becomes unstable.

\section{Results}

The basic idea behind this work is that the exomoons' albedo could be estimated from the flux change during occultation. High albedo values might suggest water ice on the surface \citep{verbiscer13}. With the analysis of periodic changes in the transit time and its duration \citep{kipping09a, kipping09b}, the mass of the moon can be calculated, from which the bulk density could also be estimated. The density of the moon helps in determining whether or not the moon contains significant amount of ice. For the albedo calculations Solar System moons and hypothetical bodies were considered as exomoons, in order to determine their observability during the occultation. The possible usage, problems and uncertainties of this method are described in the Section 4.

\subsection{Solar-type stars}

First, the flux change caused by the presence of different hypothetical exomoons was calculated using Eq. (3). As there might be a great variety of sizes, orbital distances and stellar luminosities, we restricted our analysis to certain number of example cases. For stellar distance 3 AU was used, which is the approximate location of the snowline in the Solar System. In the following calculations the Sun was used as central object, because future surveys and monitoring programs will probably focus on Sun-like stars to search for Earth-like exomoons \citep{kaltenegger10, peters13}. Beside solar type stars we also calculated some example cases for M dwarfs (see Section \ref{nextgen}), as they came into the focus of exoplanet related research recently (see the MEarth project, \citet{berta13}).

A few parameters of interesting examples are shown in Table \ref{MOdepth}. Beside Solar System bodies (Earth, Europa, Enceladus, Io), 'icy Earth' is introduced as a hypothetical moon. Icy Earth is considered to reflect as much light as Enceladus (almost 100\% of the incident light) and its radius equals to that of the Earth. It is important to note that Europa and Enceladus have no relevant atmospheres and thus we also consider the hypothetical 'icy Earth' to satisfy this criterion as well. Io does not contain water ice on the surface, but still has moderately elevated albedo because of fresh sulphur containing material. Although there is no Earth-sized moon in the Solar System, a natural satellite with this size might be considered realistic, and can serve as a useful for comparison to other objects \citep{ogihara12}. As it can be seen from Table \ref{MOdepth}, the required photometric precision is many orders of magnitude beyond the capability of current technology and such measurements will not be feasible in the foreseeable future.

\begin{table}
	\caption{Calculated values of the MO depth for different bodies around a Sun-like star. $A_\mathrm{g}$: geometric albedo, $R_\mathrm{m}$: radius of the exomoon. The distance from the central star is 3 AU in each case. The geometric albedo values are obtained from \citet{pang81, verbiscer07, verbiscer13}.}
	\centering
    \label{MOdepth}
	\begin{tabular}{c c c c}
		\hline\hline
		Name & $A_\mathrm{g}$ & $R_\mathrm{m}$ [km] & $y_3 - y_4$ [ppm] \\
		\hline
		Enceladus & 1.38 & 252 & $4.3 \cdot 10^{-7}$ \\
		Europa & 1.02 & 1561 & $1.2 \cdot 10^{-5}$ \\
		Io & 0.74 & 1821 & $1.2 \cdot 10^{-5}$ \\
		Earth & 0.37 & 6371 & $7.5 \cdot 10^{-5}$ \\
		icy Earth & 1.38 & 6371 & $2.8 \cdot 10^{-4}$ \\
		\hline
	\end{tabular}
\end{table}

\subsection{Moons in systems of M dwarfs}

\subsubsection{Different wavelengths} \label{wavelength}

It is well known that M stars are much brighter in the infrared than in the visible spectrum. For this reason it seems obvious to measure the flux at longer wavelengths, however, water ice is most reflective in the visible. In this section we discuss which wavelengths are optimal to use for successful measurements of icy moons.

\begin{figure*}
	\centering
	\includegraphics[width=120mm]{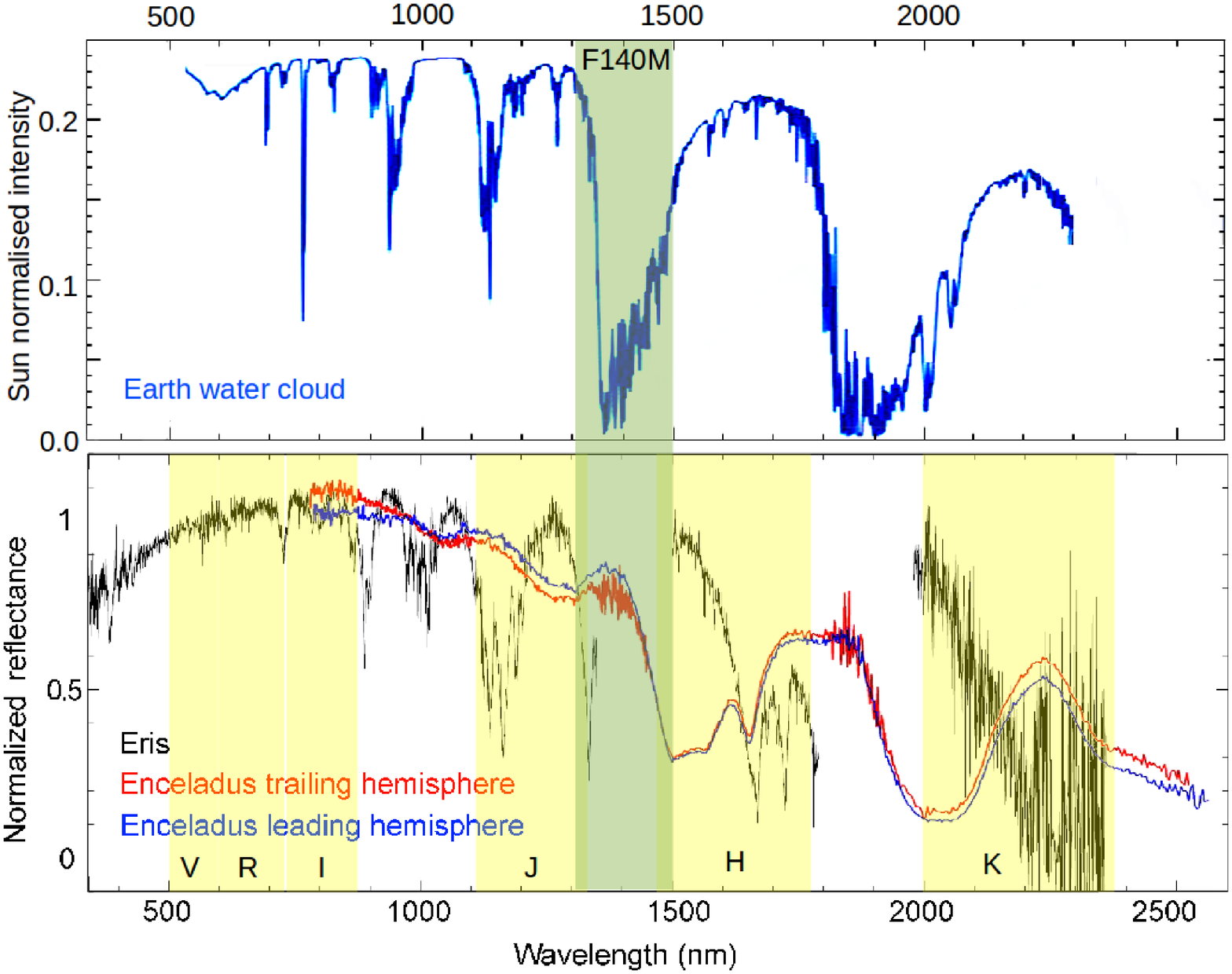}
	\caption{Top panel: Spectrum of water clouds on Earth. Bottom panel: Spectra of Enceladus (both the leading and the trailing hemispheres are indicated with blue and red colours, respectively) and Eris (black) from \citet{verbiscer06} and \citet{alvarez11}, respectively. The visible spectrum of Enceladus is not shown, but the geometric albedo is known to be 1.375 \citep{verbiscer07}. Yellow vertical bands indicate the full widths at half maximum of the V, R, I, J, H and K photometric bands. The green vertical band indicates the F140M filter band where the cloud spectrum is low but the water ice spectrum is relatively high.}
	\label{spectra}
\end{figure*}

Beside water ice, methane ice is also very reflective and such bodies are also known in the Solar System that are covered at least partially with methane ice. For this reason, if the estimated albedo is high, it may be challenging to decide whether water ice or methane ice is present on the surface. To make the distinction easier, see the differences in the spectra of Enceladus and Eris in the bottom panel of Fig. \ref{spectra} (the spectra are from \citet{verbiscer06} and \citet{alvarez11}). Eris is a good example for a small body covered by methane ice, and Enceladus is used as a representation of a moon with water ice on the surface. The spectra are normalized, but these normalized values must be in good correspondence with the geometrical albedo for both bodies: the spectra of Enceladus are normalized at 0.889 $\mu$m where the geometric albedo (at phase angle ${\approx} 3^{\circ}$) of the leading and the trailing hemispheres are 1.02 and 1.06, respectively \citep{buratti98}, and the reflectance of Eris is normalized at 0.6 $\mu$m \citep{alvarez11}, and its geometric albedo in the V band is 0.96 \citep{verbiscer13} which is very close to one. From their spectra, the albedo values at different photometric bands are estimated and presented in Table \ref{bands}. All values in the table are rough estimations, which can differ for various exomoons. The albedo of Enceladus (water ice case in the table) in the V band is from \citet{verbiscer07}.

\begin{table}
	\caption{Estimated albedo values at different photometric bands, based on the spectrum of Enceladus (water ice surface, \citet{verbiscer06, verbiscer07}) and Eris (methane ice surface, \citet{alvarez11}).}
    \label{bands}
	\centering
	\begin{tabular}{c c c c c c c}
		\hline\hline
		Surface & V & R & I & J & H & K \\
		\hline
		Water ice & 1.375 & n.a. & 1.00 & 0.84 & 0.52 & 0.43 \\
		Methane ice & 0.97 & 1.02 & 0.98 & 0.65 & 0.45 & 0.42 \\
		\hline
	\end{tabular}
\end{table}

To see the albedo variations respective to wavelength, it is best to measure the flux in all photometric bands indicated in Table \ref{bands}, but it seems that the V and the J bands are the most relevant. Measuring in these two bands provide useful data, partly because of the apparent difference in the albedo for the two kinds of ice, which helps determining whether water or methane ice is present on the surface. Another reason for the V band is the maximum reflectance of water ice, and for the J band is the maximum spectral radiance of low-mass M dwarfs. The radiance of an M6 -- M7 main-sequence star in the J band is about 5 times stronger than in the V band. Because of the larger number of photons that arrive to the detector, it is easier to observe the occultation in the J band, where the albedo of water ice is still relatively high. However, the shape of the light curve will not change in different bands.

Clouds are also very reflective and for this reason it may be challenging to determine whether we see an icy surface or clouds in the atmosphere. Since water is very abundant is the universe, we assume that clouds form from H\textsubscript{2}O. The upper panel of Fig. \ref{spectra} shows the intensity of water cloud on Earth \citep[Chapter~7]{gottwald06}. In the F140M filter band (indicated by a green vertical band in the figure) the intensity of the clouds are very low while the reflectance of water ice is moderately high (see the Enceladus' spectrum in the bottom panel). This photometric band gives opportunity to separate clouds from ice.

\subsubsection{Next generation telescopes and the occultation of exomoons} \label{nextgen}

It seems easier to estimate the albedo of satellites around M dwarfs than around larger stars. This is because the fact that the lower mass the star has, the closer the snowline is located. It means that the planet-moon system can be closer to the central star without the sublimation and/or melting of the ice on the surface. If the moon is closer to the star, then the light drop caused by its presence will be more significant in the light curve. This effect favours detection. In other words, it may be easier to observe icy moons around lower mass stars in occultation, than around more massive stars. It also means, that smaller moons may also be detected in occultation around late type stars.

The capabilities of next generation telescopes are of fundamental importance in successful albedo estimations. Beside Kepler, the most relevant missions for exomoon detection are the CHaracterising ExOPlanets Satellite \citep[CHEOPS, ][]{broeg13}, the Transiting Exoplanet Survey Satellite \citep[TESS, ][]{ricker10, ricker14}, the James Webb Space Telescope \citep[JWST, ][]{clanton12}, the PLAnetary Transit and Oscillations of stars \citep[PLATO 2.0, ][]{rauer11, rauer14} and the European Extremely Large Telescope \citep[E-ELT, ][]{ramsay14}. The photon noise levels of these instruments were calculated as described in Section \ref{calcphotonnoise}. The data used for the calculations for each instrument are shown in Table \ref{instruments}. All these numbers are subject to change, since most of these missions are in the planning and testing phase. The numbers in the last column indicate references as explained in the footnote\footnote{1: \citet{ricker14}, 2: \citet{auvergne09}, 3: \citet{broeg13}, 4: \citet{benz13}, 5: \citet{rauer14}, 6: \citet{vancleve09}, 7: \citet{avila16}, 8: \citet{dressel16}, 9: JWST webpage: http://www.stsci.edu/jwst/instruments/nircam/instrumentdesign/filters/, 10: \citet{davies10}}.

\begin{table*}
	\caption{Instrument parameters for calculating the photon noise level. $A$: aperture, $Q$: quantum efficiency, $T_{campaign}$: observation period per target field for survey instruments. In most cases rough estimations were used. Reference numbers shown in the last column are explained in the footnote. In the cases of TESS and CHEOPS the quantum efficiencies are estimated based on the $Q(\lambda)$ curves provided by the manufacturers. Asterisk indicates that instead of the quantum efficiency, the system throughput is presented.}
	\centering
	\begin{tabular}{c c c c c c c c c}
		\hline\hline
		Instrument & $A$ [m] & Filter & $\lambda_f$ [nm] & $\lambda_{min}$ [nm] & $\lambda_{max}$ [nm] & $Q$ [\%] & $T_{campaign}$ & Ref. \\
		\hline
		TESS & 0.1 & -- & 800 & 600 & 1000 & 80 & 80 days & 1\\
		CoRoT & 0.27 & -- & 650 & 400 & 900 & 63 & 150 days & 2\\
		CHEOPS & 0.3 & -- & 750 & 400 & 1100 & 67 & -- & 3, 4\\
		PLATO 2.0 & 0.68 & -- & 750 & 500 & 1000 & 50* & 2 years & 5\\
		Kepler & 0.95 & -- & 660 & 420 & 900 & 50* & 4 years & 6\\
		HST ACS WFC & 2.4 & F555W & 541 & 458 & 621 & 35* & -- & 7\\
		HST WFC3 IR & 2.4 & F125W & 1250 & 1100 & 1400 & 52* & -- & 8\\
		JWST NIRCam & 6.5 & F115W & 1200 & 1000 & 1400 & 40* & -- & 9\\
		JWST NIRCam & 6.5 & F140M & 1400 & 1300 & 1500 & 45* & -- & 9\\
		E-ELT MICADO & 39 & J & 1258 & 1181 & 1335 & 60* & -- & 10\\
		\hline
	\end{tabular}
	\label{instruments}
\end{table*}

\begin{figure}
	\centering
	\includegraphics[width=80mm]{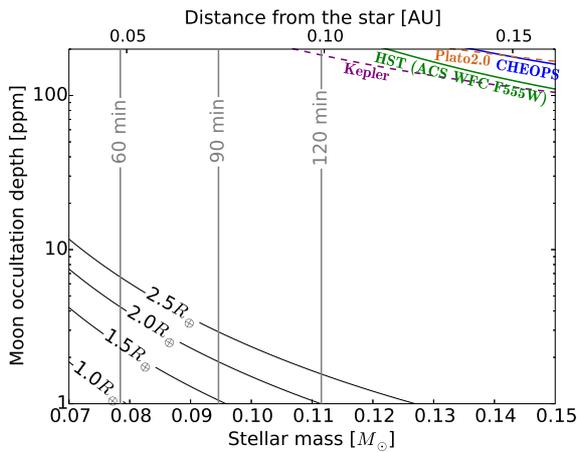}
	\caption{Radius of icy moons around different late type stars with a given MO depth. The moon's geometric albedo was set to $A_\mathrm{g}$ = 1.38 in each case (according to the reflectance of ice in the V photometric band), and the distance from the star is at the snowline, that is calculated from Eq.~\ref{snowline}, and indicated in the top axis, too. Brown, blue, green and purple curves indicate the estimated photon noise levels with 5 $\sigma$ confidence of PLATO 2.0, CHEOPS, HST and Kepler, respectively. The dashed curves for PLATO 2.0 and Kepler indicate that these are survey missions in contrast with the HST and CHEOPS observatories. Grey vertical lines indicate the 60, 90 and 120 minute MO duration contours for a moon in circular orbit around a Neptune-mass planet.}
	\label{telescopes}
\end{figure}

\begin{figure}
	\centering
	\includegraphics[width=80mm]{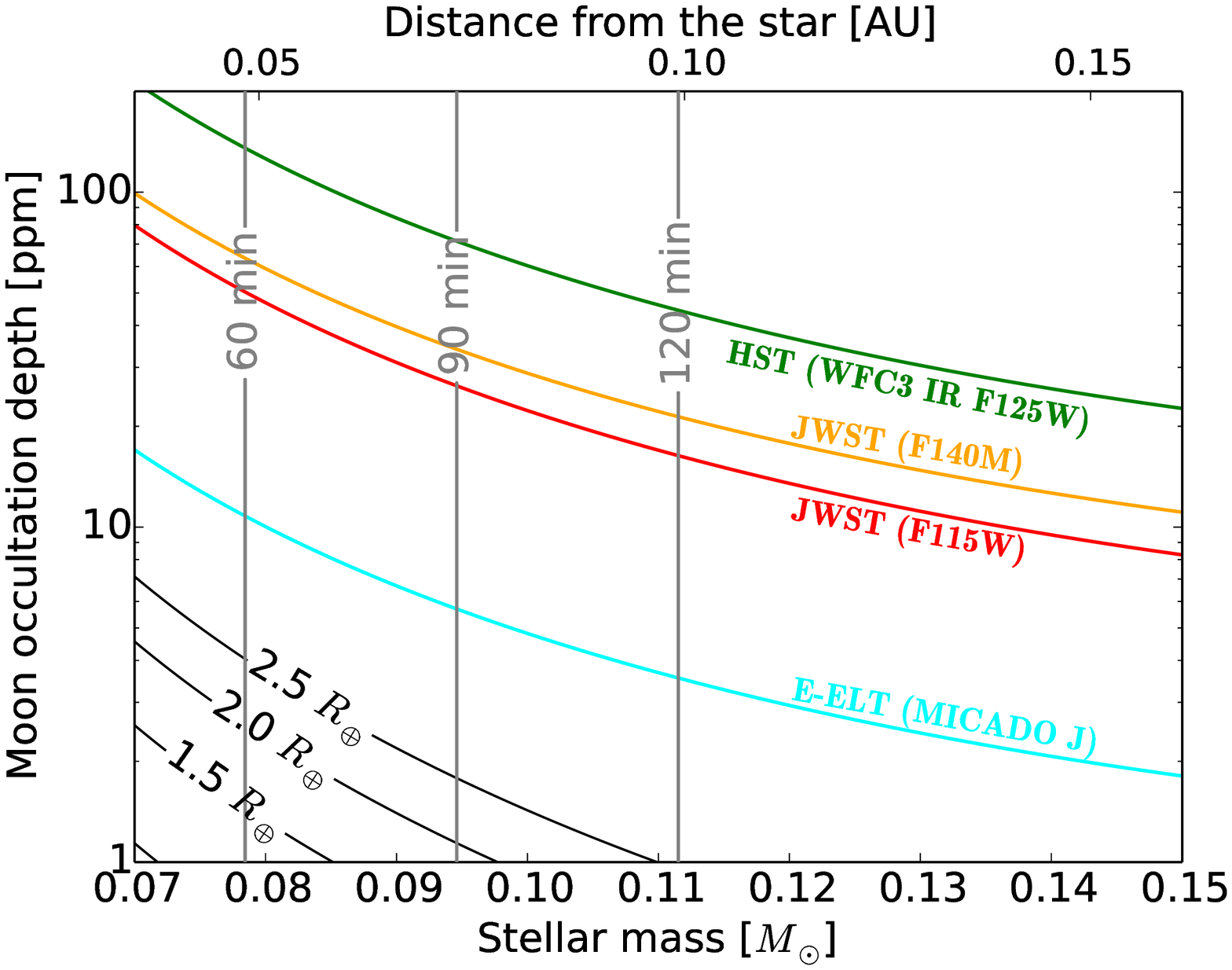}
	\caption{Radius of icy moons around different late type stars with a given MO depth. The moon's geometric albedo was set to $A_\mathrm{g}$ = 0.84 in each case (according to the reflectance of ice in the J photometric band), and the distance from the star is at the snowline, that is calculated from Eq.~\ref{snowline}, and indicated in the top axis, too. Green and light blue coloured curves indicate the estimated photon noise levels with 5 $\sigma$ confidence of HST and E-ELT, respectively. Orange and red curves depict the same for JWST but with two different filters (F140N and F115W, respectively). Grey vertical lines indicate the 60, 90 and 120 minute MO duration contours for a moon in circular orbit around a Neptune-mass planet.}
	\label{telescopes_J}
\end{figure}

Figs. \ref{telescopes} and \ref{telescopes_J} show the relation between the MO depth of icy moons and the mass of the host star in the V and J photometric bands, respectively. The objects covered in this figure are low-mass M dwarfs: M9 -- M5 according to \citet{kaltenegger09}. The MO depth is in logarithmic scale. The radius of the moon indicated by black curves was calculated for all stellar mass and MO depth pairs. For this calculation Eq. (3) was used and the albedo was fixed to that of Enceladus in each case, i.e. 1.38 for Fig. \ref{telescopes} and 0.84 for Fig. \ref{telescopes_J}. The distance from the star was set to the snowline, which was calculated from Eq.~\ref{snowline}, and is indicated in the upper axis. This is the minimum stellar distance that is required in order to prevent the melting of the surface ice. It also means that our calculation is an optimistic approach: the minimum distance and the approximately minimum moon radius are determined for a succesful detection. The vertical gray lines indicate the 60, 90 and 120 minute MO durations. In all investigated cases, the moon and the planet occults separately. For larger stellar masses the MO duration is longer, because of the larger stellar distance. The expected noise levels for the telescopes are indicated by coloured curves. Those instruments that measure in or near the J photometric band (HST WFC3 IR, JWST NIRCam and E-ELT MICADO) are shown in Fig. \ref{telescopes_J}, all the others are in Fig \ref{telescopes}, however, because of their limited capabilities, TESS, and CoRoT do not appear in Fig \ref{telescopes}.

The highest radius that is shown in Figs. \ref{telescopes} and \ref{telescopes_J} is 2.5 Earth-radii (for comparison, the radius of Neptune is 3.9 Earth-radii). We choose this 2.5 Earth-radii as an upper limit for a rocky moon with surface ice to guarantee that its mass is not greater than 5 Earth-masses. (The radius of a body with the density of Enceladus and with a mass of 5 Earth-masses is 2.58 Earth-radii.) Such large body around a Neptune-mass planet could also be called a binary planet, which would probably be a different system than the Saturn--Enceladus planet--moon pair. For simplicity, and also because we use this 2.5 Earth-radii as an upper limit, we still call them moons in the rest of the paper.

In the visible spectrum, the photon noise of the telescopes is much higher than the required precision for detecting large, rocky moons in occultation. However, in the J photometric band (see Fig. \ref{telescopes_J}) the E-ELT's photon noise is just a few ppm greater than the MO depth of large icy exomoons. (Note that the MO depth is shown in a logarthmic scale.) For example, for a 2.5 Earth-radii moon around an 0.11 stellar mass star, the difference is about 3-4 ppms, and for a 1.5 Earth-radii moon at an 0.085 stellar mass star, it's about 7 ppms. The accuracy of the measurements and additional parameters that can influence the success of the observations are discussed thoroughly in the next section.

\section{Discussion}

The findings on the observability of different exomoons, the estimation of their surface albedo, and other related considerations are summarized below.

\begin{itemize}
	\item \textit{Smaller stellar distance:} The depth of the flux drop caused by the moon in occultation as well as its observability increases with decreasing stellar distance. In case of smaller stars, the surface of the moon can keep its icy composition even for small stellar distances, because the star's luminosity is low. Small stellar distances (such as 0.01--0.1 AU in our calculations) are important regarding the long term stability of a planet-moon system, too. According to migration theories and perturbation effects, numerous criteria have to be satisfied in order to support the presence of an exomoon that survived a certain time period after it got close to the central star \citep{barnes02}. During this process they might be able to keep their satellite system protected against perturbations by the central star and other planetary objects \citep{mosqueira03}. However, the migration is not relevant, if the moon is captured or impact-ejected (which seems more likely for large satellites) after the planet's migration phase.
	\item \textit{Larger size:} Among the analysed cases, the best possibility to determine the albedo is for Earth-sized, or even larger exomoons, because the reflecting surface is larger.
	\item \textit{Higher albedo:} The higher the albedo that an exomoon has, the more stellar light it reflects, and the flux difference in the light curve is larger during the occultation. Extremely high geometric albedos (for example 1.375 in the case of Enceladus in the Solar System) are indicators for the presence of relatively young and fresh water ice since this is the only known material that could reflect such high percentage of light. The presence of water ice on the surface may imply astrobiological importance. Considering strongly reflecting atmospheric clouds, their presence also could cause high albedo. The separation of cloud and ice reflection with optical methods could be difficult, but in case of succesful measurements with an F140M filter, it may be feasible, as discussed in Section \ref{wavelength}. However, our results show that such measurements will not be achievable with the instruments presented in Section \ref{nextgen}, meaning that next generation telescopes will still not be able to separate atmospheric clouds from surface ice on exoplanetary bodies.
\end{itemize}

Beside the challenge in observation of the small flux contribution of the exomoon, analysing the measured data will also be difficult, as to fit a realistic model, because many possible configurations are needed to consider. To see the possible roles of these factors, and in order to help to plan and target future observations, in the following we evaluate how several basic parameters affect the observation and analysis.

\subsection{Accuracy}

For an adequate result from the light curve measurements, repeated observations are needed. It is not necessary that the moon is positioned exactly at the same apparent position every time when the occultation occurs. For this reason the small flux drop caused by the moon may be shorter or longer, and may not be seen at every observation. \citet{simon12} propose using an averaged light curve of several observations to deal with the scattered position of the flux change caused by the presence of the moon. They also present a four-step detection strategy for discovering exomoons by transit method. Occultation events with the same apparent positions of the moon and the planet might be rare, but in tidally locked situations, a given configuration during subsequent occultations could be frequent or regular and repeated at every revolution of the exoplanet, and the exomoon might be at the same position viewed from our line of sight.

Variable surface features of the host star, such as granules, micro-flares etc. could influence the result, producing small flux fluctuations. Such variability is expected especially in the late type stellar atmospheres related to convective motion \citep{ludwig06}. From the moon's transit measurements and from the occultation of the host planet, the expected time of the moon's occultation can be estimated, which may help in distinguishing the MO signal from the stellar surface fluctuations, especially because the occultation will cause a periodic change in the light curve.

Beside the uncertainties in flux measurements, the accuracy of albedo estimation also depends on the accuracy of the exomoon's size and the stellar distance. In our work both are assumed to be already known from transit detections, and determining their accuracy was already discussed by several authors \citep[see e.g.][]{pal08, carter08}. However, the errors from transit measurements propagate into the determination of the geometric albedo. The total error of the albedo can be obtained from
\begin{equation}
	\left( \frac{\Delta A_\mathrm{g}} {A_\mathrm{g}} \right)^2 = \frac {\Delta (y_3 - y_4)^2} {(y_3 - y_4)^2} + \frac {4 \Delta (a_\mathrm{p} / R_\star)^2} {(a_\mathrm{p} / R_\star)^2} + \frac {4 \Delta (R_\mathrm{m}/R_\star)^2} {(R_\mathrm{m}/R_\star)^2}  \, ,
    \label{error}
\end{equation}

\noindent where the three terms of the right side of the equation can safely be considered to be independent from each-other. The first term is $(1/5)^2$ since we set the measurements to $5 \sigma$. The second term is typically around (1--5\%$)^2$ \citep{pal08}. Assuming that $a_m/R_\star \geq 1$ (i.e. the moon and the planet occult separately, which was true in all cases described in Section \ref{nextgen}), and that the same number of occultations are detected for the moon as for the planet, and that these measurements were made with the same instrument, the third term can be expressed as
\begin{equation}
	\frac {\Delta (R_\mathrm{m}/R_\star)} {(R_\mathrm{m}/R_\star)} = \frac {\Delta (R_\mathrm{p}/R_\star)} {(R_\mathrm{p}/R_\star)} \cdot \frac {(R_\mathrm{p}/R_\star)^2} {Q (R_\mathrm{m}/R_\star)^2}  \, ,
\end{equation}

\noindent where $Q$ is the quantum efficiency. This error can be calculated for given geometric configurations. In any case, the last term must be smaller than the first one in Eq. \ref{error}, because the S/N ratio is much larger for transits than for occultations, hence transit measurements are much more precise. See e.g. the case of CoRot-1b which was measured in transit and occultation, as well \citep{snellen09}. The same effect is true for exomoons. It means that the first term of Eq. \ref{error} is dominating in the error of the albedo estimation.

In the calculations shown in Section \ref{nextgen} only photon noise was considered based on the quantum efficiency or on the system throughput. However, additional noise sources can also be significant, including both random noises (such as detector readout, dark current, zodiac light or airglow) and systematic noises (sky variations, tracking jitters, subpixel structure, etc.). These noise sources could even be larger in the near-infrared regime and it can be comparable to the photon noise itself for faint(er) targets. However, for brighter objects, the photometric precision is mostly limited by the photon noise and the accuracy is limited by the systematic variations. The system throughput that was considered in the calculations (even when the quantum efficiency was used, since in these cases the quantum efficiency was multiplied by $0.8$) already contains some of these extra noise sources, but in future missions these are subject to change, since their precise values will be known only after the commissioning of the instruments. In addition, we assumed 30 occultation detections for observatories that would require at least 5 years for E-ELT because the orbital period of the planet at the snowline is approximately 31 days, and the occultation can only be measured at nights. Considering that not all noise sources were included in our estimations, and that measuring 30 occultations may be unrealistic in most cases, it can be concluded that our calculations are too optimistic, and the real noise level of the instruments will be higher than presented.

According to \citet{simon07}, measuring the photometric transit timing variations ($\mathrm{TTV_p}$) leads to the estimation of the size of the exomoon. By analysing transit duration variations (TDVs) as well, rough mass estimation is also possible under favourable conditions \citep{kipping09a, kipping09b}. Using the derived mass and radius together, information on bulk density could be gained, and putting together this value with the presence of surface ice gives strong hint on the possible existence and ratio of water ice in the interior.

\subsection{Considering different parameters} \label{diffpar}

In the following we categorize parameters by their importance in influencing the results of the observations. Specific cases that might be relevant for certain exomoons, but are not usual in our Solar System are discussed below.

\begin{itemize}
\item \textit{Eccentricity:} For the analysed, relatively large exomoons we do not expect highly eccentric orbits as the large satellites in the Solar System (even the captured ones) have low eccentricity. However, perturbations might elongate the orbit, or resonances with other moons may maintain a higher eccentricity. The variable orbital velocity around the host planet influences the apparent speed of the exomoon perpendicular to our line of sight. The changing velocity influences the length of the plateau/occultation duration. Shorter plateau/MO duration means less integration time for the telescope which results in larger photon noise. The moon-star distance periodically change with the orbital phase of the moon, influencing its brightness, but this effect is very small. Eccentricity of the moon may enhance this phenomenon. It could be substantial in cases when the ratio between the stellar distance and the exomoon's semi-major axis is relatively small. \citet{sato10} described the effect of the satellite's eccentricity on the transit light curve in details, which can similarly be used to occultations with some changes.

\item \textit{Inclination:} Large tilt of the orbital plane of the exomoon relatively to the planet's orbital plane around the host star could also influence the plateau/occultation duration, especially when the exoplanet orbits at large distance from the host star. In case of large inclinations, instead of occulting, the moon may pass beside the star. The length of the plateau/occultation duration also depends on the inclination and the ratio of the exomoon's apparent distance from its host planet and the diameter of the host star. Larger stellar disk and smaller orbital distance increase the influencing effect of higher inclinations, however if the stellar disk is smaller, then the moon will not occult for high inclinations. For transits the effect of the exomoon's inclination is discussed by \citet{sato10}.

\item \textit{Reflected light from planet:} In ideal cases the exomoon's brightness might be elevated by reflected light from its host planet. The strongest reflected light flux reaches the exomoon during the 'new moon' phase (watching from the exoplanet). In different configurations mutual transits and mutual shadows affect the shape of the light curve \citep{cabrera07}. In the albedo estimation this effect of the reflected light from the planet is not significant, as the exomoon shows almost full moon phase during the occultation (looking from the Earth). This reflected light flux is negligibly small comparing to the stellar irradiation \citep{hinkel13}. The illumination from the planet with other light sources is thoroughly discussed by \citet{heller13}.

\item \textit{Orbital direction:} Retrograde orbits affect the plateau/occultation duration as stellar eclipses by the host planet are more frequent, than with a prograde orbit and can be prolonged because of the large size of giant planets \citep{forgan13}. Such eclipses may occur just before or after the occultation with the star, changing the shape of the light curve. Using the phase-folded light curve of several observations may help in filtering out such cases. Beside eclipses, it does not influence the observability whether the exomoon is approaching to or receding from the host star viewing from the Earth.

\item \textit{Effect of hemispherical asymmetry:} An exomoon that faces with different sides toward the star (and the toward the Earth) might have different brightness at different occultations, if significant albedo difference exists between the two hemispheres. Calculating with a Solar System analogy, Iapetus shows huge difference in its Bond albedo with $\sim$0.31 at the trailing, and maximum 0.1 at the leading hemisphere \citep{howett10}. The difference between the observations of the two hemispheres produces such brightness difference that can not be neglected. This effect may influence the result of the albedo estimation, but using the phase-folded light curve such flux changes may be noticed, and from the average light curve the average albedo can be estimated.
\end{itemize}

\begin{table*}
	\caption{Summary of various factors that influence the successful albedo estimation for an exomoon. The most influencing ones are listed in the left column, the less important and indifferent ones are listed in the right column.}
    \label{diffparameters}
	\centering
	\begin{tabular}{l l l}
		\hline\hline
		Substantial influence & moderate influence & minimal influence \\
		\hline
		$\bullet$ stellar distance & $\bullet$ eccentricity of the exomoon & $\bullet$ direct/retrograde orbital \\
		$\bullet$ exomoon size & $\bullet$ inclination of the exomoon & direction \\
		$\bullet$ albedo & $\bullet$ hemispherical asymmetry & $\bullet$ reflected light from the \\
		$\bullet$ star--planet size ratio & of the moon & planet \\
		$\bullet$ planetary distance & $\bullet$ mass of the planet \\
		\hline
	\end{tabular}
\end{table*}

Based on this subsection, different parameters are categorized by their importance in occultation observations. The results can be seen in Table \ref{diffparameters}. The most relevant factors influencing the observation are: 1. close stellar distance that is observationally favourable especially at small main sequence stars; 2. star/planet diameter ratio: the larger the ratio is the longer MO duration could be observed which helps in reaching smaller photon noise for the measurement (only relevant at small stars); 3. larger exomoons are easier targets, and are favourable to form by impact ejection or by capture; 4. larger planetary distance of the moon can increase the length of the plateau duration; 5. higher albedo means that the satellite is brighter, hence makes the observation easier.

\subsection{Importance of ice}

Icy moons with subsurface oceans might hold large volume of liquid water for geologic time-scales and thus may potentially be habitable or at least could be considered as favourable environments for life. In the Solar System such subsurface oceans exist in Europa, Titan, Enceladus, and possibly also in Ganymede and Callisto. The liquid phase state of these oceans are maintained by tidal heat production, freezing point depressing solved ingredients, and/or internal heat due to accretional energy conservation and radiogenic heat \citep{carr98, khurana98, kargel00, zimmer00, mccord01, schenk02, collins07, roberts08, lorenz08, postberg09, iess14, dobos15}. Having information on the existence of surface ice, using mass estimation from TTV and TDV might increase the probability to identify exomoons that are probable of having subsurface oceans.

The brightness of the ice could point to relatively young surface and active resurfacing processes. In the case of Europa the ocean gives possibility for active resurfacing thus for the existence of clean and bright ice there, although locations with non-ice ingredients can also be found on Europa, while on Enceladus the subsurface liquid water contributes to the geyser-like eruptions, and to the increase of surface albedo caused by fresh water ice crystals that fall back to the surface \citep{kadel00, fagents03, porco06, spencer06, verbiscer07, howett11}. Although other reasons could also produce elevated albedo, the very high albedo is still a strong indication of the water ice on the surface, and even of a liquid subsurface ocean.

Estimation of moons' albedo is much easier at M type stars, if the planet and the moon are close to the snowline. However, strong, regular flares and coronal mass ejections are expected, that potentially make the orbiting rocky bodies uninhabitable, because of high UV irradiation and loss of the atmosphere \citep{scalo07}. Exomoons still might be habitable because the ice cover attenuate UV radiation, and serve as a physical screening mechanism \citep{cockell98, cockell00}. The loss of the atmosphere could also easily happen at small stellar distance, but does not influence directly the oceans beneath ice sheets.

\section{Conclusions}

In this work we proposed a method for the first time that hints on the surface albedo of an exomoon and thus might indicate the existence of water ice. The argumentation presented in this work can be useful to orient future research and to identify the best candidate systems for such observations. The methods and the results of our calculations are presented firstly, then secondly those parameters are listed which substantially influence such measurements and also those that are not relevant -- as such information is useful in planning and targeting observations.

Moons of smaller stars seem to be easier to detect, if their stellar distance is smaller. In addition, the smaller the star, the closer the snowline located, meaning that rocky bodies can stay icy even if they are orbiting the star in a closer orbit (but still outside the snowline). The vicinity to the star makes the body brighter, thus an icy exomoon is easier to detect in occultation close to the snowline of an M dwarf, than at the snowline of a solar-like star. The largest flux difference is expected from large, $\sim$ 2.5 Earth-radii icy moons (about 5 Earth-mass body with similar density of Enceladus) orbiting small M dwarfs ($\sim$ 0.07--0.13 solar masses) close to the snowline. We have found that such moons cannot be observed in occultation with next generation space missions in the visual spectrum, because the instruments' photon noise is far greater than the flux drop caused by the moon. However, in the near infrared (J band), the E-ELT's photon noise is just about 5-6 ppm greater than the MO depth, despite the lower albedo of ice. This result, however, is too optimistic, because it does not take into account other noise sources, such as CCD read-out and dark signal, which become significant in the near infrared measurements. Considering that not all noise sources were included in our estimations, and that measuring 30 occultations may be unrealistic in most cases, since it would require years to achieve, it can be concluded that the real noise level of the instruments in the near infrared wavelengths will be higher than presented in Section \ref{nextgen}. Flux fluctuations caused by stellar activity makes the measurements even more difficult. We conclude that occultation measurements with next generation missions are far too challenging, even in the case of large, icy moons at the snowline of small M dwarfs. However, we expect that as future missions will further develop in the next decades, resulting in even better precisions, their noise level will be lower, hence occultations may be detected for exomoons around small M dwarfs.

We outlined the parameters that can influence the detection. Based on assumptions, the orbital and physical parameters of exomoons might be highly diverse, thus we evaluated a range in the space of several parameters. We categorized these parameters by their influence on the success of MO detections (see Table \ref{diffparameters}).

We also discussed the possible properties of the most suitable exomoons for characterization which may be useful once the instrumentation is available. Exomoons around Neptune-sized exoplanets of small stars seem to be the best targets for occultation-based albedo estimation opposite to Jupiter-like exoplanets, if the satellites, formed by impact ejection or capturing, are large enough. In case of such satellites the surface ice cover might be co-accreted if the ejection happened before the clearing up of the circumplanetary disk, or formed by ejection from planets composed mostly of water ice, like Neptune and not of hydrogen, like Jupiter. Binary-exchange capture is also a possibility for a giant planet to have a large icy satellite (Neptunes are better candidates for this process than Jupiters).

Based on our work, the first albedo estimations of exomoons from occultations are expected around low mass M dwarfs at small stellar distances. These moons will probably be large, and have high albedo which may imply the presence of ice on the surface.

\begin{acknowledgements}
We thank the useful comments for the anonymous referee who helped in substantially improving the manuscript. We also thank Ren\'e Heller for the helpful discussion. VD has been supported by the Hungarian OTKA Grant K104607, the Hungarian National Research, Development and Innovation Office (NKFIH) grant K-115709, the Lend\"ulet-2009 Young Researchers Program of the Hungarian Academy of Sciences, and the ESA PECS Contract No. 4000110889/14/NL/NDe. \'AK has been supported by the Astrophysical and Geochemical Laboratory and the COST TD 1308 project. PA has been supported by the LP2012-31 project.
\end{acknowledgements}

\bibliographystyle{aa}
\bibliography{ref}

\end{document}